# Increasing Server Availability for Overall System Security: A Preventive Maintenance Approach Based on Failure Prediction

**Ayman M. Bahaa Eldin[1], Hoda K. Mohamead[1], Sally S. Deraz[2]**

1  Computer and Systems Engineering Department, Faculty of Engineering, Ain Shams University.

2  IBM Cairo Technology Development Center. Cairo, Egypt.

**ABSTRACT**

Server Availability (SA) is an important measure of overall systems security. Important security systems rely on the availability of their hosting servers to deliver critical security services. Many of these servers offer management interface through web mainly using an Apache server. This paper investigates the increase of Server Availability by the use of Artificial Neural Networks (ANN) to predict software aging phenomenon. Several resource usage data is collected and analyzed on a typical long-running software system (a web server). A Multi-Layer Perceptron feed forward Artificial Neural Network was trained on an Apache web server dataset to predict future server resource exhaustion through uni-variate time series forecasting. The results were benchmarked against those obtained from nonparametric statistical techniques, parametric time series models and empirical modeling techniques reported in the literature.

## 1. INTRODUCTION

It has been observed that software applications executing continuously over a long period of time, such as web servers, show a degraded performance and increasing rate of failures [1]. This phenomenon has been called 'software aging' [2]. This may be due to memory leaks, unreleased file-locks and round-off errors. Currently, researchers have been looking for methods to counteract this phenomenon by what is so called 'software rejuvenation' methods [3] such as applying a form of 'preventive' maintenance. This could be done by, for example, occasionally stopping the software application, cleaning its internal state and then restarting [3] to prevent unexpected future system outages. This allows for scheduled downtime at the discretion of the user, which suggests an optimal timing of software rejuvenation.

The goal of this research is to use prediction of resources exhaustion as well as further information for deriving the best rejuvenation schedule.

## 2. RELATED WORK

The software aging problem is currently approached either by building analytical models for system degradation such as probability models, linear and nonlinear statistical models, expert systems and fractal base models [4], or by empirically studying the software systems based on measurements. Reported results focus on applying statistical regression techniques to monitored numerical system data, and few attempts were reported on the use of Wavelet Networks in software aging [10,11].

In [1], they studied the development of resource usage (i.e. response time and memory usage) of an Apache web server while subjecting it to an artificial workload. Among the system parameters, they monitored free physical memory, used swap space, which follows a seasonal pattern, and response time of the Apache web server. Non-parametric statistical techniques and parametric time series models were employed for analyzing the collected data.

They also focused on how to model seasonal patterns, and determine the model order and showed how the exploitation of the seasonal variation can help in adequately predicting the future resource usage.

Since their predication of any parameter in time series analysis was based on the previous values of the same parameter, they proposed multivariate time series models to be employed for investigating the interactions between various system resources as a future enhancement. They also proposed studying the influence of the configuration of the operating system, and the server on the aging behavior in details as another future enhancement.

In [4], a best practice guide for building empirical models for the prediction of performance variables of computing systems was proposed. They validated its applicability in three case studies; by modeling and forecasting Apache web server response time, and free available memory, as well as the call availability of a telecommunication system.

Their primary goal was to gain better understanding of the way the system works. Based on observations about its past behavior, statistical models were derived to capture its dynamics, and to make forecasts about its future behavior.

Steps of best practice guide are:

1. System Observation: Building time teries from system log files.
2. Variable selection: where cross benchmark of three variable selection techniques to filter out the most predictive variables; forward selection, backward elimination, and probabilistic wrapper (PWA). Then Root-mean-square-errors (RMSE) is calculated for the three cross benchmarked variable selection techniques. They selected the variable set yielding the smallest RMSE on the validation data set.
3. Model Building: Variable sets identified in the previous step to cross benchmark four modeling techniques for short and long term predictions are used. These modeling techniques are multivariate linear regression (ML), support vector machines (SVM), radial basis functions (RBF), and universal basis functions (UBF).
4. Sensitivity Analysis: investigating each variable contribution to the overall quality of the model, by evaluating potentially nonlinear relationships between a variable and the target variable. They took each of the four models built in the previous model building step and removed each of the variables found in the variable selection step one at a time. They then calculated the change in model error RMSE for the validation, and the generalization data set.

It is interesting to note that variables to which the model is highly sensitive are not necessarily highly correlated with the target.

Based on the best practice guide they proposed, they recommended focusing the modeling efforts on the variable selection step because it yields consistently the largest improvement (one order of magnitude) in prediction performance in their test cases. Results from sensitivity analysis, they suggested that variable selection with a nonlinear wrapper approach may improve model quality further. However favoring nonlinear over linear multivariate regression techniques yields to consistent improvements, but they may not always be significant.

They also proposed searching for causality inference (i.e. finding root cause) and getting optimal reaction schemes in an effort to close the loop as further investigation.

From the previous literature review, time series models could be fitted to the data collected to help predicting the future resource usage. Attributes subject to software aging are monitored and related data is collected aiming at predicting the expected exhaustion of resources like real memory and swap space. Then, non-parametric statistical techniques and parametric time series models are employed to analyze the collected data and estimate time to exhaustion via extrapolation for each resource, usually assuming linear functions of time.

ANN is a relatively simple model that can produce adequate forecasts and can efficiently deal with seasonality in resource usage and system performance.

In this work, ANNs are used to forecast Swap Space Usage, Free Physical Memory and Response Time of an apache web server and results are cross benchmarked against reported results in the literature based on parametric and nonparametric statistical analysis, and empirical modeling techniques.

## 3. COLLECTED DATA

Data and results reported in [1] and [4] are used to carry on further analysis using an Artificial Neural Network approach. The collected data is from a Linux web server with an artificial load approaching its maximum optimal load level.

The setup that was used for collecting the data consisted of a server running Apache version 1.3.14 on a Linux platform, and a client connected via an Ethernet local area network. The fact that the Linux OS stores an abundance of system information in the /proc virtual file system was exploited. For example, the file /proc/meminfo contains information about the usage of physical memory and swap space, while information on the system load can be found in the file /proc/loadavg. From the /proc file system and with the help of the Linux monitoring tool "procmon", measurements were periodically collected. "httperf" were used to generate requests with constant time intervals between two requests.

Each request accesses one of five specified files of sizes 500 bytes, 5 kB, 50 kB, 500 kB, and 5MB on the server. "httperf" is not only a workload generator, but it can also be employed for monitoring performance information. The measurements provided include the reply rate (i.e., the number of responses received from the server per unit time), the response time (i.e., the interval from the time "httperf" sends out the first byte of a request until it receives the first byte of reply), and the number of timeout errors (i.e., the total number of requests for which no response was received from the server due to timeout errors). For collecting resource usage data over a long time period, a shell program was used to run "httperf" periodically.

As for the connection rate, a value of 400 requests per second was chosen, which puts the web server in an overload state, and should speed up software aging.

Among the system parameters of the web server monitored during a period of more than 3.5 weeks are the response time of the web server, the used swap space, and the free physical memory. The three time series are shown in Figure 2, Figure 6 and Figure 8 respectively. Data were collected during experiments in which the web server was put in a near overload condition indicating the presence of software aging.

## 4. NEURAL NETWORK APPROACH

### 4.1. Artificial Neural Networks

ANN is a class of flexible nonlinear models that can discover patterns adaptively from the data. Given an appropriate number of nonlinear processing units, neural networks can learn from experience and estimate any complex functional relationship with high accuracy.

Numerous successful ANN applications have been reported in the literature in a variety of fields including pattern recognition and forecasting. For a comprehensive overview of ANN the reader is referred to [5].

### 4.2. ANN for time series analysis

The usage of ANN for time series analysis relies entirely on the data that were observed and is powerful enough to represent any form of time series. ANN can learn even in the case of noisy data and can represent nonlinear time series. For example, Given a series of values of the variable x at time step t and at past time steps x(t), x(t-1), x(t-2),…, x(t-m), we look for an unknown function F such that; x(t+n)=F[x(t), x(t-1), x(t-2),…, x(t-m)], which gives an n-step predictor of order m for the quantity x.

Although many types of neural network models have been proposed, the most popular one for time series forecasting is the Multi-Layer Perceptron (MLP) feed forward model [6]. A multi-layer feed forward network with at least one hidden layer and a sufficient number of hidden neurons is capable of approximating any measurable function [7]. A feed-forward network can map a finite time sequence into the value that the sequence will have at some point in the future [8].

Feed forward ANNs are intrinsically non-linear, nonparametric approximators, which makes them suitable for complex prediction tasks. The ANN sees the time series X1,…,Xn in the form of many mappings of an input vector to an output value [9]. The time-lagged values x(t), x(t-1), x(t-2),…,x(t-m) are fed as inputs to the network, which once trained on many input-output pairs, gives as output the predicted value for yet unseen x values.

### 4.3. ANN for software aging forecasting

In software aging, we do not have a well defined model describing the aging process that one would like to study. All that is available are measurements of the variables of interest (i.e. time series). Therefore, we propose, in this work, an artificial neural network approach for the problem of forecasting the attributes, subject to software aging that are monitored, with the objective of predicting the expected exhaustion of resources like real memory and swap space.

### 4.4. Proposed neural network structure

For our problem, we decided to use a fully connected, MLP, feed forward ANN with one hidden layer, and the back propagation learning algorithm [8]. The input nodes are the previous lagged observations while the output node is the forecast for the future values.

Hidden nodes with appropriate non-linear transfer functions (activation functions) are used to process the information received by the input nodes.

### 4.5. ANN topology and parameter selection

The number of ANN input neurons determine the number of periods the neural network "look into the past" when predicting the future. Whereas it has been shown that one hidden layer is sufficient to approximate continuous function [5], the number of hidden units necessary is not known in general. To examine the distribution of the ANN main parameters (i.e. number of time lags and number of neurons in the hidden layer), we conducted a number of experiments, where these parameters were systematically changed to explore their effect on the forecasting capabilities. These estimations of the network's most important parameters although rough, allowed us to choose reasonable parameters for our ANN.

Multi Layer Perceptron (MLP) Feed Forward Neural networks were constructed as illustrated in Figure 1, for the three parameters under study.

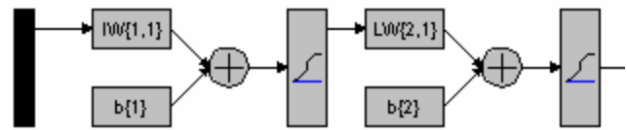

Figure 1. Implemented MLP neural network

We divided the collected data into three segments, one to train the ANN, one for validation, and the third for testing. The testing segment is used to evaluate the forecasting performance of the ANN in predicting the performance parameters' values.

To measure the accuracy of the training and forecasting, the difference between the expected values from the ANN and the actual values of the test set is calculated. Three error measures are calculated, namely Root Mean Square Error (RMSE) Mean Absolute Percent Error (MAPE), and Symmetric Mean Absolute Percent Error (SMAPE), Although RMSE is enough to assess the results, other error measures are calculated to compare the results with others

Increasing Server Availability for Overall System Security: A Preventive Maintenance Approach Based on Failure Prediction

*4.5.1. Mean Absolute Percent Error (MAPE).*

MAPE is calculated by averaging the percentage difference between the fitted (forecast) line and the original data:

$$MAPE = \frac{1}{n}\sum_{t=1}^{n} 100 \left|\frac{e_t}{y_t}\right|$$

Where y represents the original series and e represents the original series minus the forecast, and n the number of observations.

*4.5.2. Symmetric Mean Absolute Percent Error (SMAPE).*

SMAPE calculates the symmetric absolute error in percent between the actual X and the forecast F across all observations t of the test set of size n. The formula is:

$$SMAPE = \frac{1}{n}\sum_{t=1}^{n} 100 \frac{|X_t - F_t|}{0.5(X_t + F_t)}$$

**4.6. Forecasting the exhaustion of the Apache server resources**

We use the ANN described above and the data introduced in [1] and [4] to predict three Apache performance variables; response time, free physical memory and swap space used, in order to obtain predictions about possible ***near-future*** failures due to resource exhaustion.

This dataset was split into three segments; the first segment is used to train the ANN and the second segment is used to tune the ANN parameters (i.e. number of time lags and number of neurons in the hidden layer) and validation. The third segment is used to measure the ANN generalization performance on data which has not been presented to the NN during parameter tuning.

*4.6.1. Forecasting response time for Apache server.*

The dataset was collected in a 7-day period with a connection rate of 350 per second. Response time is defined as the interval from the time a client sends the first byte of a request until the client receives the first byte of the reply. Figure 2 illustrates a plot of the measured response time for the Apache server. The increasing response time over time has been explained previously by resource exhaustion [1].

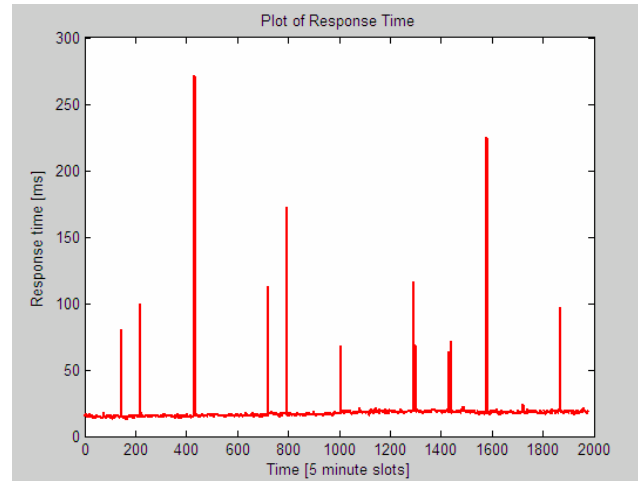

Figure 2: Response Time

We notice significant spikes in the plot that could be due to operating system's page faults or memory access. Since our objective is not to predict such spikes; rather, our intention is to predict the pattern in response time, we performed data preprocessing to filter out such spikes. To do so, we have preprocessed the data by calculating the median of the response time for the next 24 hours over a sliding window. In Figure 3, we show a plot of the preprocessed response time with vertical solid line segmenting the data into the set used in training and validation, and the set used in testing.

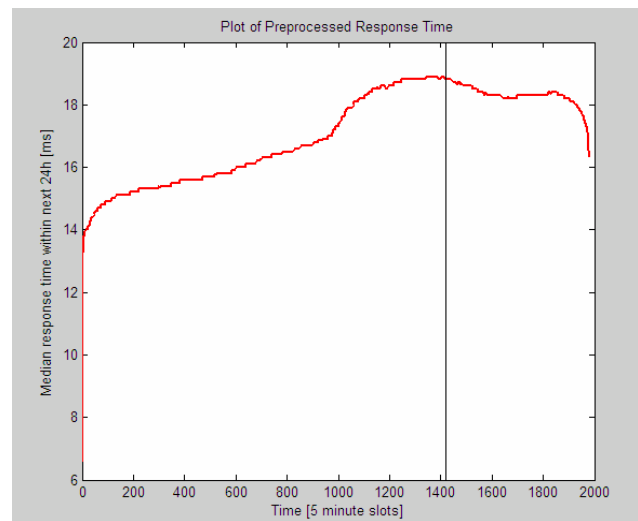

Figure 3: Response Time Preprocessed

If we can accurately forecast response time, then we could trigger a preventive maintenance actions based on this forecast.

One major problem for using an ANN for time series analysis and forecasting is the determination of the network parameters, mainly the number of neurons in the hidden layer, and the window of past time, or the number of time lags, used to forecast future values. The approach used here is to select different combinations of these 2 parameters and each time the ANN is trained with the training set and tested using the test set. In each trial, the RMSE is calculated and the combination that produce the lowest value is used for the construction of the ANN predictor. Those experiments are summarized in Table 1.

**Table 1: RMSE for different number of lags and hidden layer neurons**

| No. of neurons<br>No. of Time Lags | 2 | 3 | 4 |
|---|---|---|---|
| 3 | 0.0405 | 0.0259 | 0.026 |
| 4 | 0.0392 | 0.0256 | 0.02466 |
| 5 | 0.0405 | 0.02698 | 0.02473 |
| 6 | 0.17798 | 0.01347 | 0.00408 |
| 7 | 0.0494 | 0.03047 | 0.0278 |
| No. of neurons<br>No. of Time Lags | 5 | 6 | 7 |
| 3 | 0.02382 | 0.0405 | 0.0265 |
| 4 | 0.02536 | 0.0192 | 0.0238 |
| 5 | 0.0244 | 0.03121 | 0.02464 |
| 6 | 0.004028 | 0.00715 | **0.00169** |
| 7 | 0.0278 | 0.02598 | 0.0331 |

Figure 4 shows a graph for RMSE values for the number of time lags and number of neurons in the hidden layer. The number of time lags and hidden layer neurons that give the best accuracy is highlighted by a circle.

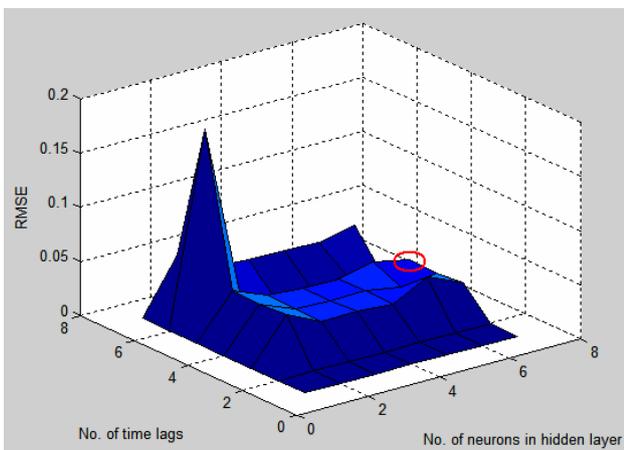

Figure 4: Conducted experiments for response time

Table 2 summarizes the RMSE, MAPE and SMAPE for the response time of the Apache server for the testing dataset. As shown in Figure 5, we were able to forecast the response time accurately using the MLP described in Figure 1 with 6 input neurons (time lags), 7 neurons in the hidden layer and a sigmoid nonlinear transfer function.

**Table 2: Response Time forecasting accuracy**

| Error measures for predicted data | SMAPE | MAPE | RMSE |
|---|---|---|---|
| ANN approach | 0.00775% | 0.0109% | 0.00169 |

As shown in Figure 5 the results obtained are accurate.

Based on RMSE, comparing with the results reported in [4] using universal basis functions, multivariate linear regression, support vector machines and radial basis functions empirical modeling techniques, we find that our proposal provide less RMSE than the mentioned techniques and hence it is considered better than those techniuques.

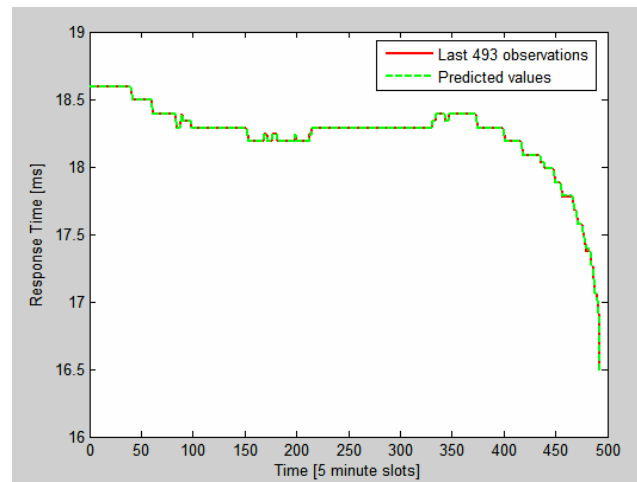

Figure 5: Response Time forecasting results

### 4.6.2. Forecasting swap space used for Apache server.

This dataset was collected on a 25-day period with connection rate of 400 per second.

Figure 6 shows swap space usage for the Apache server. It is clear that it follows a seasonal pattern and that considerable increases in used swap space occur at fixed intervals.

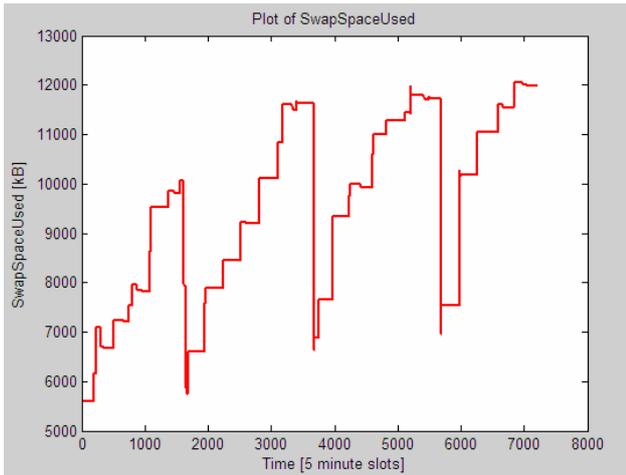

Figure 6: Swap Space Used

Table 3 shows the RMSE, MAPE and SMAPE for the forecasts of the Swap Space Used of the Apache server for the testing dataset using the MLP described in Figure 1 with 3 input neurons (time lags), 4 neurons in the hidden layer and a sigmoid nonlinear transfer function.

As seen in Table 3, the results obtained by the ANN approach are far more accurate than the results obtained by the non-parametric statistical approach reported in [1].

**Table 3: Swap Space Used evaluation**

| Error measures for the predicted data | SMAPE | MAPE | RMSE |
|---|---|---|---|
| Non-Parametric Statistical approach | 4.313% | 4.47% | 612.46 |
| ANN approach | 0.292% | 0.295% | 113.189 |

In Figure 7, we show a plot of the last 3208 observations of the measured SwapSpaceUsed (the testing dataset) and the predicted values obtained by the ANN approach, which shows accurate predictions.

### 4.6.3. Forecasting free physical memory for Apache server.

The dataset was collected in a 7-day period with a connection rate of 350 per second.

Figure 8 shows a plot over time of the free physical memory of the Apache server with vertical solid line segmenting the data into the set used in training and validation, and the set used in testing.

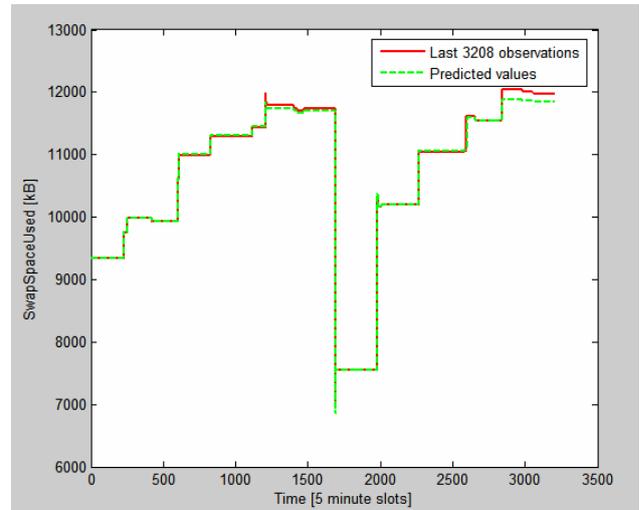

Figure 7: Swap Space Used forecasting results

Our goal is to forecast future values of this variable.

We notice that as physical memory approaches its minimum allowed lower limit, the system frees up memory by paging, giving an irregular utilization pattern.

We have normalized the data by scaling it to the interval [0,1].

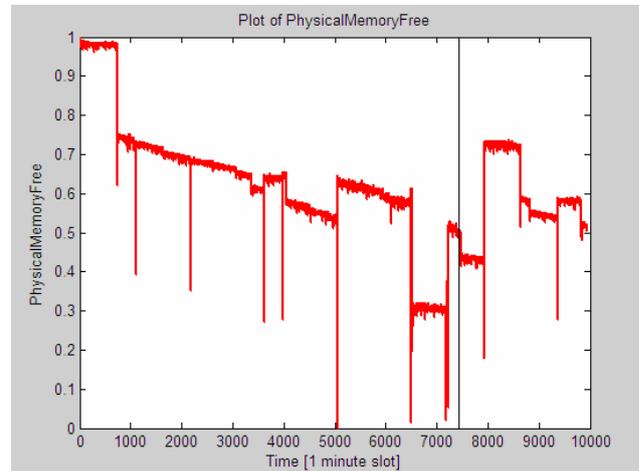

Figure 8: Physical Free Memory

Table 4 shows the RMSE, MAPE and SMAPE for the Free Physical Memory of the Apache server for the testing dataset forecasts using the MLP described in Figure 1 with 4 input neurons (time lags), 2 neurons in the hidden layer and a sigmoid nonlinear transfer function.

Table 4: Physical Free Memory forecasting accuracy

| Error measures for the predicted data | SMAPE | MAPE | RMSE |
|---|---|---|---|
| ANN approach | 1.078% | 1.093% | 0.0117 |

In Figure 9, we show a plot of the last 2483 observations of the measured physical free memory (the testing dataset) and the predicted values obtained by the ANN approach, which shows accurate forecasts.

Based on RMSE the obtained results are more accurate than the results reported in [4] obtained from universal basis functions, multivariate linear regression, support vector machines and radial basis functions empirical modeling techniques.

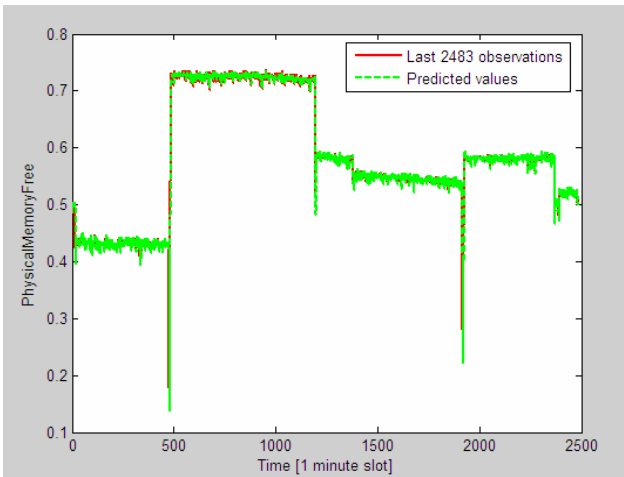

Figure 9: Physical Free Memory forecasting results

## 5. CONCLUSION AND FUTURE WORK

In this work we have forecasted the development of response time, memory usage and swap space used, that are related to software aging, of an Apache web server subjected to a synthetic load for 25 days, using a proposed Artificial Neural Networks approach. We showed that a feed forward Artificial Neural Networks is able to accurately predict the future behavior of these performance variables.

The results obtained were benchmarked against those reported in the literature that are based on parametric and non-parametric statistical techniques and empirical modeling techniques and were more accurate.

Future work involves extending the proposed Artificial Neural Network approach to empirically model the process of software aging in order to attempt to define an optimal software rejuvenation policy. Also Sensitivity Analysis for the selected resource usage parameters is to be carried to select the best set.

## 6. ACKNOWLEDGEMENT

The authors would like to thank Professor Kishor S. Trivedi, the Hudson Chair in the Department of Electrical and Computer Engineering at Duke University, for valuable discussions during the development of this work and for his review of the manuscript and Michael Grottke for providing the Apache performance dataset.

## 7. REFRENCES

[1] Grottke M., L. Lie and K. Vaidyanathan, K. Trivedi: Analysis of software aging in a web server, IEEE Transactions on Reliability, Vol. 55, No. 3, September 2006, pp.411-420, 2006

[2] Dohi T., K. Goseva-Popstojanova and K. S. Trivedi, Analysis of software Cost Models with Rejuvenation, Proc. Of the IEEE Intl. Symp. On High Assurance Systems Engineering, HASE, 2000.

[3] Huang Y., C. Kintala, N. Kolettis and N. Fulton, Software Rejuvenation: Analysis, Module and applications, In proc. Of the 25th IEEE Intl. Symp. On Fault Tolerant com-puting (FTCS-25), Pasadena, CA, 1995

[4] Guenther A. Hoffmann, Kishor S. Trivedi, and Miroslaw Malek: A Best Practice Guide to Resource Fore casting for Computing Systems, IEEE Transactions on Reliability, VOL. X, NO. X, DECEMBER 2007

[5] Hornik, K., Stinchcombe, M., White, H., Multilayer feedforward networks are universal approximators. Neural Networks 3, 551-560, 1989

[6] Zhang, G. Peter and Qi, Min, Neural network forecasting for seasonal and trend time series, European Journal of Operational Research 160, 501-514, 2005

Increasing Server Availability for Overall System Security: A Preventive Maintenance Approach Based on Failure Prediction

[7] Siegelmann H. and Sontag Eduardo D., Neural nets are universal computing devices. Technical Report SYSCON-91-08, Rugters Center for Systems and Control, 1991

[8] Hassoun, M. H., Fundamentals of Artificial Neural Networks, MIT Press, 1995

[9] Chakraborty, K., Mehrota, K., Chilukuri, K. Mohan and Ranka, S., Forecasting the behaviour of multivariate time series using neural networks., Neural Networks 5, 961-970, 1992

[10] Xu, J., You, J. and Zhang, K., A Neural-Wavelet based Methodology for software Aging Forecasting, IEEE International Conference on Systems, Man and Cybernetics, Volume 1, Issue , 10-12 Oct. 2005 Page(s): 59 - 63 Vol. 1, 2005.

[11] Ning, M. H., Yong Q., Di, H., Ying, Ch. And Zhong, Z. J., Software Aging Prediction Model Based on Fuzzy Wavelet Network with Adaptive Genetic Algorithm, Proceedings of the 18th IEEE International Conference on Tools with Artificial Intelligence (ICTAI'06), Pages 659-666, 2006.

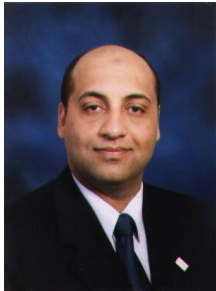

**Dr. Ayman Mohammad Bahaa Eldin**
Assistant Professor, Computer and Systems Eng. Dept. Ain Shams University
Received his B.SC. , Master of Science, PhD. In Computer Engineering from Ain Shams University in 1995, 1999, and 2004 respectively. His field of research are in the area of cryptography, computer sand network security. Ayman.bahh@eng.asu.edu.eg

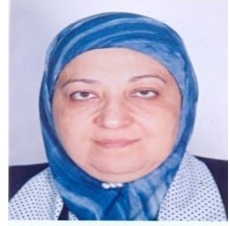

Dr. Hoda Korashy is an associative prof. at Ain shams university, Faculty of Engineering, computers & systems department. Field of interest: database and data mining, intelligent systems and agents.

**Sally Sobhy Derz**

Research Engineer in IBM Cairo, Egypt.
She received her B.SC and Master of Science in computer Engineering from Ain Shams University
Her Research interests are in Data Mining

**ملخص البحث**

يعتبر مدى توفر أجهزة الخوادم وقدرتها على تنفيذ المهام الموكلة لها من أهم معايير قياس الأمان الكلي لنظام المعلومات. هناك العديد من الأنظمة الأمنية التي تعتمد على توفر الأجهزة الخادمة للقيام بعملها وعادة ما يتم إدارة نظم التأمين تلك عن طريق واجهة تطبيق ويب تعتمد في الأغلب على نظام الأباتشي Apache. في هذا البحث يتم دراسة زيادة توفر أجهزة الخوادم عن طريق استخدام الشبكات العصبية الاصطناعية للتنبؤ بظاهرة تقادم البرمجيات. ومن أجل ذلك يتم تجميع عدد كبير من بيانات استهلاك موارد خادم يعمل لفترة طويلة ثم يتم تحليلها. تم استخدام شبكة عصبية ذات تغذية أمامية متعددة المراحل وتدريبها على البيانات المجمعة من خادم أباتشي لتوقع الاستهلاك المستقبلي المفرط للموارد عن طريق تحليل السلسلة الزمنية. كما تم مقارنة النتائج التي حصلنا عليها مع الطرق الأخرى المنشورة سابقا.